\documentclass{article}
\usepackage{spconf,amsmath,graphicx,hyperref,xcolor}
\usepackage{caption}       
\usepackage{tipa}          

\title{Improving Language Identification for Multilingual Speakers}

\name{
    Andrew Titus\textsuperscript{$\star$}\thanks{Thanks to Russ Webb, Man-Hung Siu and Stephen Shum for reviewing.},
    Jan Silovsky\textsuperscript{$\star$},
    Nanxin Chen\textsuperscript{$\star \dagger$}\thanks{Work performed by Nanxin Chen while interning at Apple.},
    Roger Hsiao\textsuperscript{$\star$},
    Mary Young\textsuperscript{$\star$},
    Arnab Ghoshal\textsuperscript{$\star$}
}

\address{
    \textsuperscript{$\star$}Apple, \textsuperscript{$\dagger$}Johns Hopkins University\\
    \texttt{\{\href{mailto:andrew\_titus@apple.com}{andrew\_titus},
              \href{mailto:jsilovsky@apple.com}{jsilovsky},
              \href{mailto:rhsiao@apple.com}{rhsiao},
              \href{mailto:mary\_young@apple.com}{mary\_young},
              \href{mailto:aghoshal@apple.com}{aghoshal}\}@apple.com},\\
    \texttt{\href{mailto:bobchennan@jhu.edu}{bobchennan@jhu.edu}}
}

\begin{document}

\maketitle

\begin{abstract}
Spoken language identification (LID) technologies have improved in recent years from
discriminating largely distinct languages to discriminating highly similar languages or even
dialects of the same language. One aspect that has been mostly neglected, however, is discrimination
of languages for multilingual speakers, despite being a primary target audience of many
systems that utilize LID technologies. As we show in this work, LID systems can have a high
average accuracy for most combinations of languages while greatly underperforming for others
when accented speech is present. We address this by using coarser-grained targets for the acoustic
LID model and integrating its outputs with interaction context signals in a context-aware model to
tailor the system to each user. This combined system achieves an average $97\%$ accuracy across
all language combinations while improving worst-case accuracy by over $60\%$ relative to our baseline.
\end{abstract}


\section{Introduction}
\label{sec:intro}

Automatic speech recognition (ASR) systems are becoming increasingly ubiquitous in today's world as
more and more mobile devices, home appliances and automobiles add ASR capabilities. Although many
improvements have been made in multi-dialect \cite{Yoo19, Padi19}, multi-accent \cite{Fukuda18, Jain18}
and even truly multilingual \cite{Mueller18, Bagchi19, Hori19} ASR in recent years, they often only
support a small subset of languages \cite{Li19}. In order to get a satisfactory Word
Error Rate (WER) for a larger range of languages, language identification (LID) models
have been combined with monolingual ASR systems to allow utterance-level switching for a
larger set of languages \cite{Gonzalez14} with reasonable accuracy, even over a set of up to
8 candidate languages.

Supporting many dozens of distinct languages, however, can lead to both low LID accuracy and
high computational load from running many recognizers in parallel. This is even further complicated
by the existence of recognizers for multiple locales (language-location pairs) for
the same language, such as American English and British English. Fortunately, only a small
percentage of people can speak more than three languages fluently \cite{EC12}, thus greatly
constraining the range of possible classifications for a given user. In the case of bilingual
speakers, LID is as simple as a binary classification.

Such prior knowledge has been incorporated into LID systems to improve accuracy. In \cite{Wan19},
the tuplemax loss function was introduced to incorporate prior knowledge of installed dictation locales
directly into the training of the LID model and achieved nearly a 40\% relative
improvement in LID accuracy over their baseline (cross-entropy loss) system, even with
79 candidate locales. However, there are three practical limitations to their analysis as it
relates to common LID use cases. The first is that the majority of both the training and evaluation
data in \cite{Wan19} was collected from monolingual speakers, creating a mismatch with the conditions
under which the LID system would be used. This is in fact quite common in the LID literature, as even
standard benchmarks such as the NIST Language Recognition Evaluation (LRE) \cite{LRE17} do not
currently make such a distinction in the datasets they use. The second limitation is that the system
was allowed to run on long utterances with overlapping windows and no latency constraints, an
impractical assumption for a deployed dictation system. Finally, accuracy is reported as an
average of pairwise language identification tasks without taking into account the relative
frequency of the language pairs within the user population.

In this paper, we demonstrate how our method of incorporating prior knowledge about usage patterns
into our LID system for dictation allows us to make highly accurate decisions for multilingual speakers
across a space of over 60 locales \cite{Apple19} while keeping latency low. We then present
in-depth error analysis methods for on-device language ID systems, including a novel metric,
Average User Accuracy (AUA), that leverages statistical information about the frequency of
installed dictation locales to better capture the expected experience across a population of
users than previous metrics.

\section{System overview}
\label{sec:system_overview}

Our LID system is composed of two stages: an \textit{acoustic model} and a
\textit{context-aware model}. The former is responsible for making predictions based
on the evidence encompassed in the speech signal, whereas the latter is responsible
for making final predictions by integrating the posteriors produced by the
acoustic model with assorted interaction context signals. These context signals encompass
information about the conditions
under which the dictation request was made, including information about installed dictation locales,
the currently selected dictation locale and whether the user toggled the dictation locale
before making the request. The context information is also essential in the situation when
the speech signal is too short for the acoustic model to produce a reliable prediction
(\textit{e.g.}, short ambiguous utterances such as \textipa{/naIn/}, which could be the negative
``nein'' in German or the number ``nine'' in English if the user has both English and German installed).
The predictions provided by this two-stage LID system are then used to select the correct monolingual
automatic speech recognition (ASR) model for the given request, similar to the system described in
\cite{Gonzalez14}.

\subsection{Acoustic model}
\label{subsec:acoustic_model}

Our acoustic model uses standard 40-dimensional Mel-filter bank features as input.
The features are extracted with the standard window length of 25ms and window shift of 10ms with
mean normalization applied afterwards. Before any feature vectors are processed by the acoustic
model, the audio signal is pre-processed by a speech activity detector to avoid processing audio
that is mainly non-speech. The acoustic LID model starts processing audio once a minimum
speech duration threshold is met. 

The targets of the acoustic model were initially the locales themselves. However, it can be quite
difficult to train the model to discriminate between locales due to two primary reasons. The
first is that the ground truth labels for locale can be noisy due to users picking locales
based primarily on location, rather than accent or dialect (\textit{e.g.}, expatriates
using the locale of their current country, rather than their native country). The second
is that some locales can be quite similar acoustically (\textit{e.g.}, American English
and Canadian English) and differ primarily in other non-acoustic areas, such as the language
model. Thus, we experiment with other choices of targets for the acoustic model by
clustering locales together into target classes (see Section~\ref{subsec:am_training_strategies})
based on language and similarities with other locales for the same language.

Regardless of the choice of target class during training, we want the acoustic model to predict
language posteriors during inference. This allows the context-aware model described in
Section~\ref{subsec:context_aware_model} to better handle situations like those
described above where a user may speak a given language $l$ in a style consistent with one locale $\ell_a$
of language $l$, but only have a different locale $\ell_b$ of language $l$ installed. 
Therefore, while we may train with separate classes for locales of the same language (see
Sections~\ref{subsec:am_training_strategies} and \ref{subsec:res_worst_case}), we simplify the
determination of language during inference by taking the maximum of all logits from $y$ that
map to each language.

We also investigated various neural network architectures in our preliminary experiments, including variants of
LSTMs \cite{Marchi18}, CNN-BLSTMs \cite{Cai19}, models with self-attention \cite{Zhu18}, etc.
Even after hyperparameter tuning, we did not observe any particular architecture to be substantially better
than another. A key component of all the well-performing models, including the LSTM-based ones, turned out
to be temporal pooling layers \cite{Snyder16}, a key feature of state-of-the-art
X-vector LID systems \cite{Snyder18}. Our final model, summarized in Table~\ref{table:am_architecture},
consists of a CNN frontend and applies mean and standard deviation temporal pooling.
The total number of trainable parameters is 8M.

\begin{table}[h!]
\centering
\begin{tabular}{ |c|c|c|c| } 
 \hline
 Layer type & Layer setup & \# of layers \\ 
 \hline
 Convolutional  & 128 filters, 15x4 kernels & 4  \\
 & 1x2 max pooling  &  \\ 
\hline
 Fully connected & 1024 nodes & 4 \\ 
\hline
 Fully connected & 512 nodes & 1 \\ 
\hline
 Temporal pooling & Mean, StdDev & 1 \\ 
\hline
 Fully connected & 1024 nodes & 1 \\ 
\hline
 Softmax & Acoustic model targets & 1 \\ 
 \hline
\end{tabular}
\caption{Acoustic model architecture.}
\label{table:am_architecture}
\end{table}

\vspace*{-\baselineskip}

\subsection{Context-aware model}
\label{subsec:context_aware_model}

The context-aware model serves two main goals. The first is to leverage aforementioned context
under which the dictation request was made to improve the recognition accuracy. The second goal
is to resolve locale posteriors from the language posteriors produced by the acoustic model during
inference. To accomplish these goals, we chose a simple architecture: first, we project language
posteriors to locale posteriors by setting each locale $\ell$ posterior to the its
corresponding language $l$ posterior. Next, we mask out posteriors for all locales that are not
installed on the device and re-scale the remaining posteriors to sum to 1. Finally, we integrate
these masked posteriors with the context signals in a na\"{i}ve Bayes model trained using statistics
gathered from internal users during the development process. The na\"{i}ve Bayes model was chosen
because it allowed us to reason about the conditional probabilities being used in the model without
requiring a large amount of data. This interpretability allowed us to easily debug issues and tune
thresholds for the model during the development process as more data became available.
Additionally, although the independence assumption does not hold for some features (for example, currently
selected dictation locale is indeed dependent upon the list of installed dictation locales), na\"{i}ve
Bayes classifiers work surprisingly well under such conditions \cite{Rish01}.

\section{System evaluation}
\label{sec:system_evaluation}

\subsection{Average User Accuracy}
\label{subsec:aua}

As mentioned in Section~\ref{sec:intro}, most speakers can only
dictate in up to three languages. The overall accuracy across all supported locales
is thus not representative of the target application. Accuracy assessed for smaller tuples
of locales (\textit{e.g.}, pairs and triples) reflects the nature of the target application
much more closely. However, for the sake of model comparison, a scalar representation of a
model's overall performance is more useful. One such scalar representation is the average
of these tuple-wise accuracies. This is still not representative of the real world usage
patterns, though, because the frequency of observing specific locale tuples depends on the
size of the population speaking the given languages. 

To address this, we developed a custom metric which we refer to as Average User Accuracy (AUA).
AUA allows us to better reflect population-level usage patterns in model evaluation and
thus get a better intuition behind what the average user's experience is like while using the LID system. 
We compute AUA as a weighted average of the accuracies for the top-$N$ most frequently used locale tuples,
where the weights are proportional to the average number of monthly users for the given tuple.
To compute locale tuple accuracy for a tuple $T$ containing $m$ locales $\ell_1, \ell_2 \dots \ell_m$,
we first run the LID system on only utterances that have all locales in $T$ installed (thus restricting
samples to multilingual speakers, a key differentation from previous work) and compare the predicted
locale for each utterance to the correct ground truth locale. Because there can be some imbalance
in number of utterances available for each ground truth locale within $T$, we compute the locale
tuple accuracy for $T$ as an unweighted average of the accuracies for subsets of utterances corresponding
to each ground truth locale, rather than the overall accuracy over all utterances combined.

\subsection{Worst-case performance}
\label{subsec:worst_case_performance}

Besides the AUA which serves as our primary metric, special attention is paid to the worst-case
performance of the model. We focus on the aforementioned accuracies specific to the ground truth
locales within each locale tuple because there could be significant differences in these accuracies
due to the model being biased towards one locale or another within a given tuple. An extreme case
of this was illustrated by analysis of
the en-IN (Indian English) and hi-Latn (Hindi transliterated to Latin characters) locale tuple,
where the model was heavily biased towards Hindi despite having a reasonable average score.
Analyzing this pair helped to expose difficulties with accented speech in the en-IN locale (as
well as other locales) that were previously undiscovered and led us to the decision to model some locales
for the same language as separate targets during training (see Section \ref{subsec:acoustic_model}).

\section{Results}
\label{sec:results}

We report results for an internal corpus composed of 128k dictation utterances from strictly
multilingual speakers with corresponding interaction context information. 
The AUA weights are based on average number of unique monthly users for each locale tuple and is
restricted to the top-100 most commonly used locale tuples.
A fixed window size of 2 seconds of audio (starting after speech activity is detected) is used during
decoding in all reported experiments with the exception of Section~\ref{subsec:incremental_inference},
which deals with flexible window sizes. Additionally, to get a fairer comparison of acoustic models
in Sections~\ref{subsec:am_training_strategies} and \ref{subsec:res_worst_case}, we set all context
signals except for the list of installed dictation locales to zeros (\textit{i.e.}, to ignore the
signal) in order to minimize the effect of the context-aware model and instead simply mask out
the posteriors for the uninstalled dictation locales.

\subsection{Acoustic Model Training Strategies}
\label{subsec:am_training_strategies}

We initially compared two strategies of training the model to predict language posteriors
using locale-annotated training data. In the first case ($M_{\text{locales}}$), we train the
model with locales as targets and then max-pool locale posteriors to language posteriors
as described in Section~\ref{subsec:acoustic_model}.
In the second case ($M_{\text{langs}}$), we combine data from all locales associated with a
particular language together before training and then train the model with languages as targets. Because the
distribution of the training data is not completely even, neither across languages nor
across locales, we employ class-specific cross-entropy weights during training. We conclude based on the results
presented in Table~\ref{table:am_comparison} that $M_{\text{langs}}$ yields better results in terms of AUA.

\subsection{Improving worst-case performance}
\label{subsec:res_worst_case}

Analysis of worst-case performance revealed that en-IN was successfully
classified by $M_{\text{langs}}$ in only $38\%$ of trials for the (en-IN, hi-Latn) locale tuple.
The performance of hi-Latn in the same locale tuple was $81\%$, leading to an unweighted average of
$59.5\%$. Despite this locale tuple accuracy being low, the weight of this locale tuple is not high
enough to sway the overall AUA, highlighting the importance of conducting worst-case performance analysis.

We observed that the en-IN was underrepresented in
the pooled English training data. With the intent of keeping the number of training samples
for all classes balanced, we decided to model en-IN as a separate language class $M_{\text{en-IN}}$.
Training $M_{\text{en-IN}}$ in this way improved en-IN recognition accuracy to $66\%$ without any
meaningful change in AUA. Further, we hypothesized that because of large variability of the
dialects and accents present in the English class, the model could benefit even more
by separating English into more fine-grained language classes than just en-IN. Hence, in our final model
$M_{\text{en-IN + L2}}$, we defined three separate classes denoted as en-L1, en-L2 and en-IN,
where en-L1 is composed of locales where English is natively spoken and en-L2 is composed of locales
where English is spoken as a second language (excluding en-IN). This composition of the training
data led to significant improvements in worst-case performance for multiple locales as
shown in Fig.~\ref{fig:am_comparison} without significantly affecting AUA.

\begin{table}[h!]
    \centering
    \begin{tabular}{ |c|c|c|c|c| } 
        \hline
        Model & $M_{\text{locales}}$ & $M_{\text{langs}}$ & $M_{\text{en-IN}}$ & $M_{\text{en-IN + L2}}$ \\
        \hline
        AUA & $92.2\%$ & $92.9\%$ & $92.8\%$ & $92.5\%$ \\ 
        Worst-case & $45.9\%$ & $38.3\%$ & $58.5\%$ & $66.1\%$ \\ 
        \hline
    \end{tabular}
    \setlength{\abovecaptionskip}{1pt}
    \setlength{\belowcaptionskip}{1pt}
    \caption{Comparison of acoustic model training strategies}
    \label{table:am_comparison}
\end{table}

\begin{figure}
    \centering
    \includegraphics[scale=0.67]{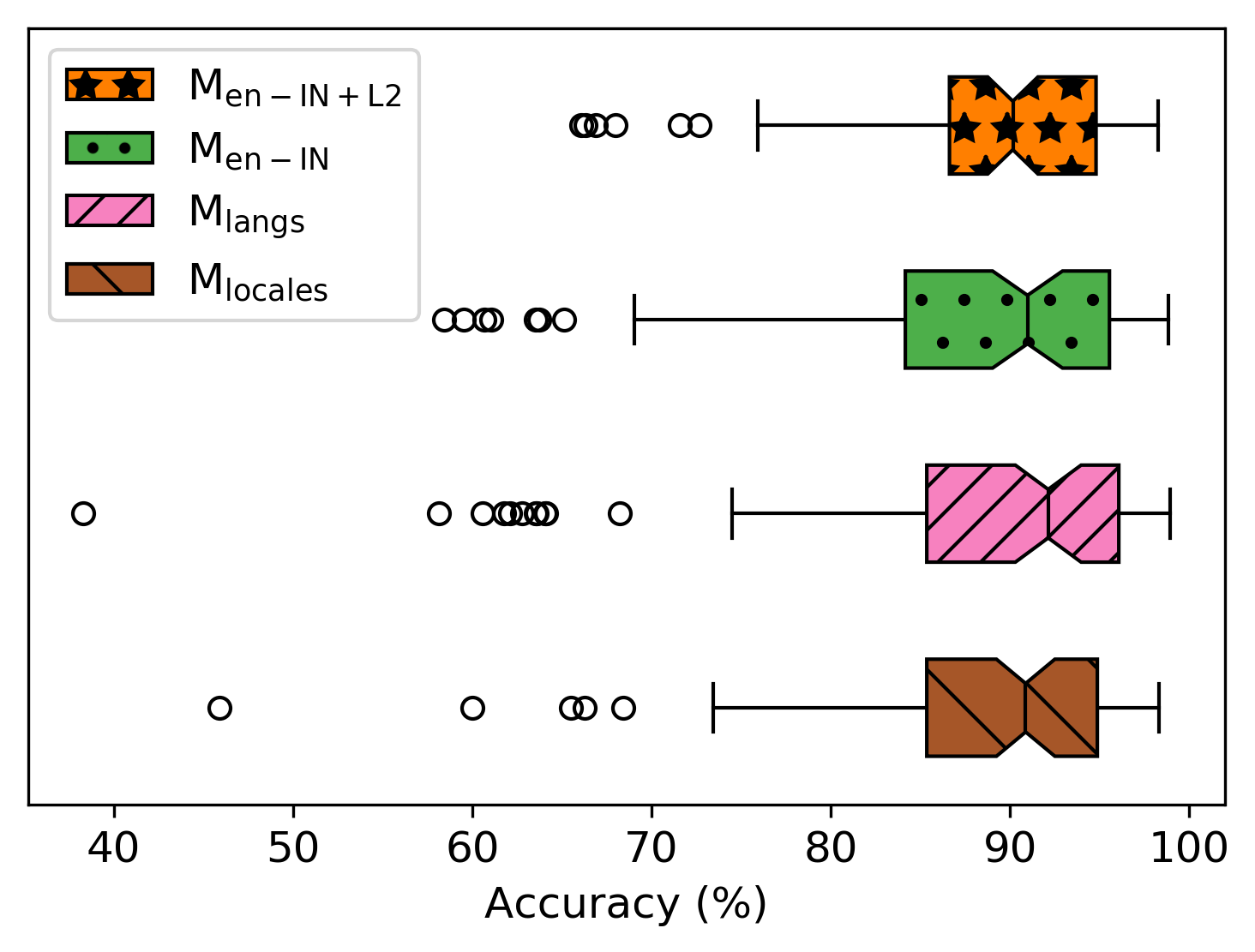}
    \setlength{\abovecaptionskip}{1pt}
    \setlength{\belowcaptionskip}{1pt}
    \caption{Locale tuple accuracies for acoustic model}
    \label{fig:am_comparison}
\end{figure}

\vspace*{-\baselineskip}

\subsection{Effect of Context-Aware Modeling}
\label{subsec:ca_results}

\begin{figure}
    \centering
    \includegraphics[scale=0.67]{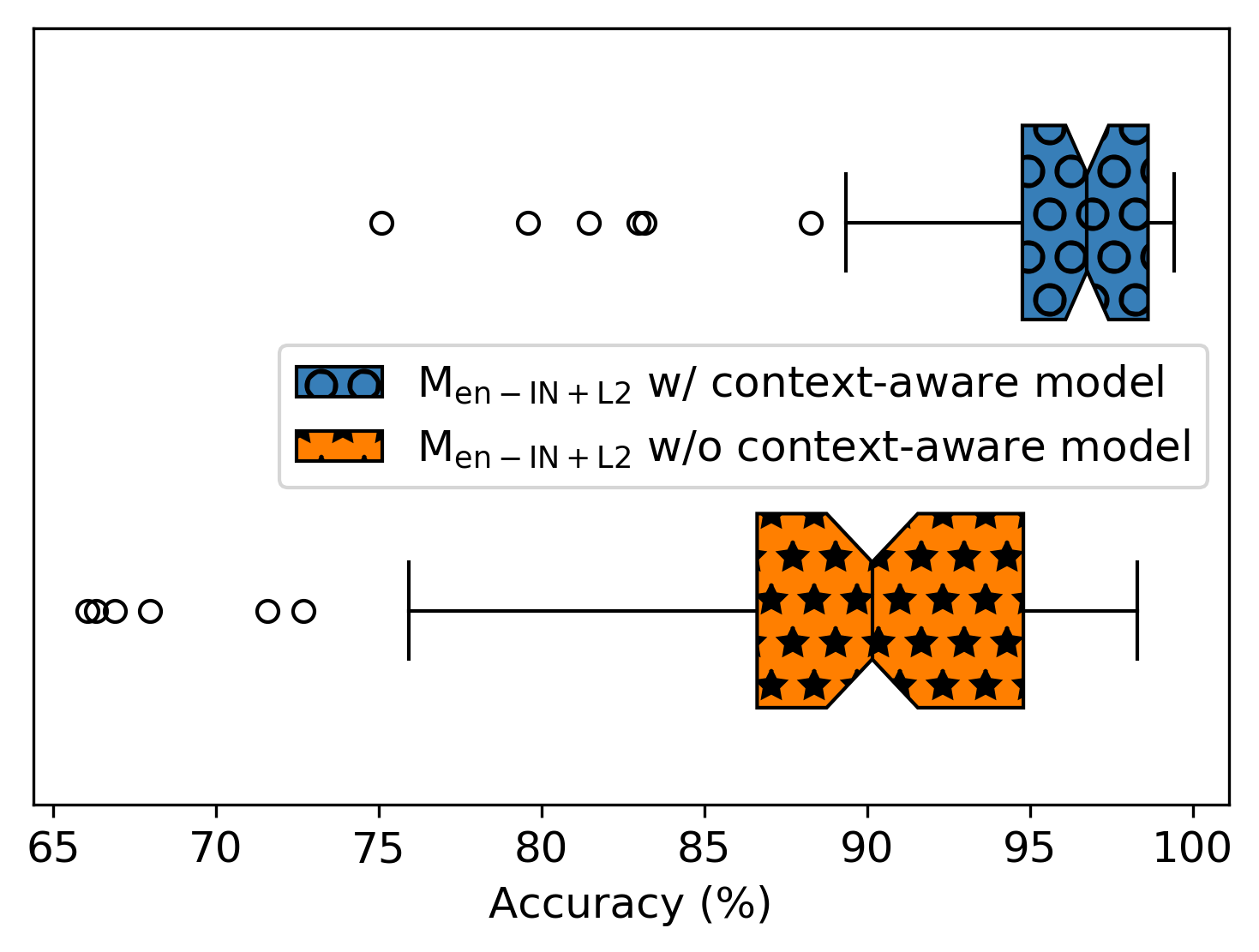}
    \setlength{\abovecaptionskip}{1pt}
    \setlength{\belowcaptionskip}{1pt}
    \caption{Locale tuple accuracies for context-aware model}
    \label{fig:ca_improvements}
\end{figure}

Incorporating the context-aware model does more than simply mask out posteriors for locales not installed
on the user's device. The model also takes into account the currently selected dictation locale and whether
the user toggled to this locale directly before making the request. These features are helpful in situations
where the acoustic model is not particularly confident in any of the locales, as well as when the
user has multiple locales installed that map to the same language (for example, hi-IN and hi-Latn are
both Hindi, but use the Devanagari and transliterated Latin scripts, respectively). By utilizing the context-aware
model to incorporate the context signals, we improve AUA from $92.5\%$ to $97.0\%$ while also improving
the worst-case locale tuple accuracy from $66.1\%$ to $75.1\%$ (see Figure~\ref{fig:ca_improvements}).

\vspace*{-\baselineskip}

\subsection{Incremental Inference}
\label{subsec:incremental_inference}

One benefit of the temporal pooling layer mentioned in Section~\ref{subsec:acoustic_model} is that
allows for variable-sized input during decoding. We conducted experiments to see how well the combined LID
system generalizes on short inputs under 2 seconds in length and found that we can get highly accurate
results in some cases even with only 1 second of speech. We used this fact to reduce average latency
using the following strategy:

\begin{itemize}
    \setlength\itemsep{0em}     
    \item{Run LID system on $t_{\text{min}}$ seconds of audio.}
    \item{If maximum posterior below confidence threshold $c$, run again on $(t_{\text{min}} + t_{\text{interval}})$ seconds of audio.}
    \item{Continue until maximum posterior exceeds $c$ or we hit $t_{\text{max}}$ seconds of audio.}
\end{itemize}

After tuning these parameters to balance accuracy and latency with the computational load of running
the model on-device, we found that we could reduce average latency from 2 seconds to 1.2 seconds
without reducing AUA by more than $0.05\%$ absolute.
The intuition behind the lack of AUA degradation is that most utterances that exceed $c$ for a given
locale $\ell$ after only a short amount of audio (\textit{e.g.}, $t_{\text{min}}$ seconds) are still
classified as $\ell$ when more audio context is given. This is consistent with the robustness of temporal
pooling layers to short audio segments demonstrated in \cite{Snyder16}. By reducing latency in this
manner, we not only improve the user experience, but also reduce the computational load of running
multiple recognizers by stopping the recognizers for all other languages besides the detected one
early in the request.

\section{Conclusion}
\label{sec:conclusion}

In this work, we present a set of modeling and analysis techniques for improving the performance
of spoken language identification systems for multilingual speakers. We do this by incorporating
prior knowledge about the usage patterns of such speakers into both the training and evaluation
of language ID systems to improve both average and worst-case performance. By using these techniques,
we achieve our final model that achieves $97\%$ AUA with less than 2 seconds of audio on average,
all while keeping worst-case accuracy for multilingual speakers above $75\%$.

\clearpage        
\bibliographystyle{IEEEbib}
\nocite{*}
\bibliography{refs}

\end{document}